\documentclass[12pt]{article}
\def\newpic#1{}

\input{epsf}

\newtheorem{theorem}{Theorem}
\newtheorem{lemma}{Lemma}

\newtheorem{prob}{Problem}

\newenvironment{proof}{{\bf Proof.}}{\hfill\rule{2mm}{2mm}}

\newtheorem{remarka}{Remark}

\def\NP {{\bf NP}}

\def\mod { {\rm mod }  }
\def\mod#1{ {\ (\rm mod \ }#1)}

\newtheorem{prelem}{{\bf Theorem}}

\newenvironment{lem}{\begin{prelem}{\hspace{-0.5
               em}{\bf.}}}{\end{prelem}}

\newtheorem{prelemm}{{\bf Lemma}}

\newenvironment{lemm}{\begin{prelemm}{\hspace{-0.5
               em}{\bf.}}}{\end{prelemm}}

\title{\bf On the Complexity of the Circular Chromatic Number }
\author{
{\bf H. Hatami and R. Tusserkani} \\
\\
{\small\it Department of Computer Engineering} \\
Sharif University of Technology \\
P.O. Box 11365--9517, Tehran, I.R. Iran }

\date{}
\begin{document}
\maketitle
\begin{abstract}
Circular chromatic number, $\chi_c$ is a natural generalization
of chromatic number. It is known that it is \NP-hard to
determine whether or not an arbitrary graph $G$ satisfies $\chi(G)
= \chi_c(G)$. In this paper we prove that this problem is \NP-hard
even if the chromatic number of the graph is known. This answers a 
question of Xuding Zhu. Also we prove
that for all positive integers $k \ge 2$ and $n \ge 3$, for a
given graph $G$ with $\chi(G)=n$, it is \NP-complete to verify if
$\chi_c(G) \le n- \frac{1}{k}$.
\end{abstract}

\section{Introduction}
We follow~\cite{West} for terminology and notation not defined
here, and we consider finite undirected simple graphs. Given a
graph $G$, an edge $e=xy$ of $G$ and a triple $(H;a,b)$ where $a$
and $b$ are distinct vertices of the graph $H$, by {\sf replacing
the edge $e$ by $(H;a,b)$,} we mean taking the disjoint union of
$G-e$ and $H$, and identifying $x$ with $a$ and $y$ with $b$. For our purposes,
it does not matter whether $x$ is identified with $a$ or with $b$.

For two positive integers $p$ and $q$, a {\sf $(p,q)$-coloring} of
a graph $G$ is a vertex coloring $c$ of $G$ with colors
$\{0,1,2,\ldots,p-1\}$ such that
$$(x,y) \in E(G) \Longrightarrow q \le |c(x)-c(y)| \le p-q.$$
The {\sf circular chromatic number} is defined as
\begin{center}
$\chi_c(G)=$ inf $\{p/q : G$ is $(p,q)$-circular colorable$\}.$
\end{center}
So for a positive integer $k$, a $(k,1)$-coloring of a graph $G$
is just an ordinary $k$-coloring of $G$. The circular chromatic
number  of a graph was introduced by Vince~\cite{Vince} as ``the
star-chromatic number'' in 1988. He proved that for every finite graph
$G$, the infimum in the definition of the circular chromatic
number is attained, so the circular chromatic number $\chi_c(G)$
is always rational. He also proved, among other things, that
$\chi-1 < \chi_c \le \chi$, and $\chi_c(K_n)=n$.

For a $(p,q)$-coloring $\phi$ of a graph $G$, let $D_{\phi}(G)$
be the digraph with vertex set $V(G)$ and for every edge $xy$ in
$G$ there is a directed edge $(x,y)$ in $D_{\phi}(G)$, if
$\phi(y) - \phi(x) = q \mod{p}.$
\begin{lemm}
{\rm \cite{Guichard}} For a graph $G$, $\chi_c(G) < p/q$ if and
only if $D_c(G)$ is acyclic for some $(p,q)$-coloring $c$ of $G$.
\end{lemm}

 The question determining which graphs have
$\chi_c=\chi$ was raised by Vince~\cite{Vince}. It was shown by
Guichard~\cite{Guichard} that it is \NP-hard to determine whether
or not an arbitrary graph $G$ satisfies $\chi_c(G)=\chi(G).$
In~\cite{Zhu} X. Zhu surveyed many results on circular
chromatic number and posed some open problems on this topic, among
them the following problem (\cite{Zhu}, Question 8.23).

\begin{prob}
What is the complexity of determining whether or not
$\chi_c(G)=\chi(G)$, if the chromatic number $\chi(G)$ is known?
\end{prob}

We answer this question, using the following theorem.
\begin{lem}
{\rm \cite{KLS}} It is \NP-hard to determine whether a graph is
$3$-colorable or any coloring of it requires at least $5$ colors.
\end{lem}

\section{Complexity}

Consider the graph $K^-$ which is obtained from a copy of $K_4$
with vertices $v_1$, $v_2$, $v_3$, and $v_4$, by removing the
edge $\{v_1, v_2\}$. In the following trivial lemma all
equalities are in $\mathcal{Z}_4$.
\begin{lemma}
\label{$K^-$}
 In every $(4,1)$-coloring $c$ of $K^-$,

\begin{itemize}
\item[(a)]  if $c(v_1)=c(v_2)$, then $D_c(K^-)$ is acyclic and
has no directed path between $v_1$ and $v_2$.
\item[(b)] if $c(v_1)-c(v_2)=1$, then $D_c(K^-)$ is acyclic and
has a directed path from $v_1$ to $v_2$.
\item[(c)] if $c(v_1)-c(v_2)=2$, then $D_c(K^-)$ has a cycle.
\end{itemize}
\end{lemma}

\begin{figure}[ht]
\input{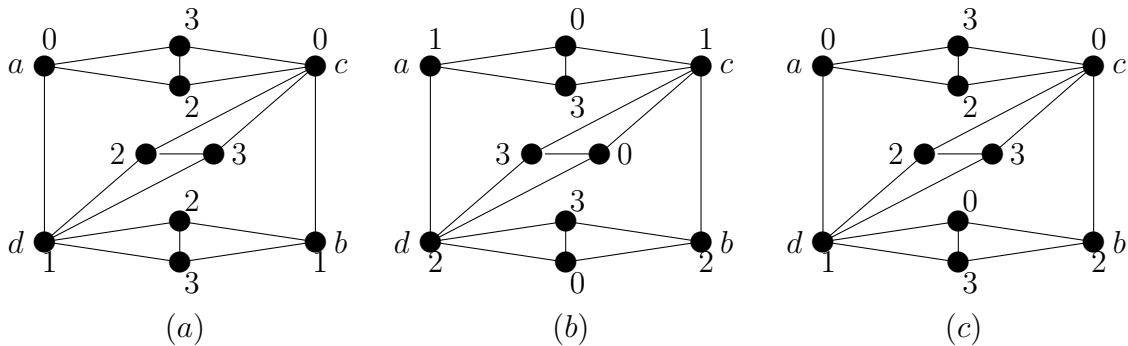}
\caption{\label{graphh}The graph $H$ and its desired colorings}
\end{figure}

Consider the graph $H$ shown in Figure~\ref{graphh}. One can
easily check that $\chi(H)=4$ and we have the following Lemma.

\newpage
\begin{lemma}
\label{H} Consider the graph $H$ shown in Figure~\ref{graphh}.

\begin{itemize}
\item[(a)] For every $(4,1)$-coloring $c$ of $H$,
 if $c(a)=c(b)$, then $D_c(H)$ has a cycle.
\item[(b)] For every $0 \le x < y \le 2$, there is a coloring $c$ for
$H$ such that $c(a)=x$, $c(b)=y$, and $D_c(H)$ is acyclic and has
no directed path from $b$ to $a$.
\end{itemize}
\end{lemma}
\begin{proof}
{\bf (a)} Without loss of generality assume that $c(a)=c(b)=0$.
For all cases except when $c(c)=c(d)=1$ and $c(c)=c(d)=3$, one
can easily check  by Lemma~\ref{$K^-$}(c) that $D_c(H)$ has a
cycle. Without loss of generality assume that $c(c)=c(d)=1$. Now
by Lemma~\ref{$K^-$}(b) there are directed paths from $d$ to $b$,
$b$ to $c$, $c$ to $a$ and $a$ to $d$. Thus $D_c(H)$ has a cycle.

\noindent {\bf (b)} Such colorings are given in
Figures~\ref{graphh}(a),~\ref{graphh}(b),~\ref{graphh}(c).
\end{proof}

\begin{theorem}
Given a graph $G$ and its chromatic number, the problem of
determining whether or not $\chi_c(G)=\chi(G)$ is \NP-hard.
\end{theorem}

\begin{proof}
For every graph $G'$, we construct a graph $G$ such that $\chi(G)=4$, and if $G'$ is
$3$-colorable, then $\chi_c(G)<4$, and if $G'$ is not
$4$-colorable, then $\chi_c(G)=4$. Thus by Theorem~A the result
is proven.

Construct the graph $G$ by replacing every edge of $G'$ by
$(H;a,b)$. Obviously, for every nontrivial graph $G'$,
$\chi(G)=4$.

First suppose that $G'$ is $3$-colorable. So we can properly color the
vertices of $G'$ with $0$, $1$, and $2$. Now by Lemma~\ref{H}(b),
this coloring can be expanded to a $(4,1)$-coloring $c$ of $G$
such that in $D_c(G)$ the copies of $H$ are acyclic, and also for
every two vertices $u$ and $v$ of $G'$, there is no path from $u$
to $v$ in $D_c(G)$ if $c(u) > c(v)$. This implies that $D_c(G)$
is acyclic. So $\chi_c(G)<4$.

Next suppose that $G'$ is not $4$-colorable. So in any
$(4,1)$-coloring $c$ of $G$ there are two adjacent vertices $u$
and $v$ of $G$ such that $c(u)=c(v)$. So by Lemma~\ref{H}(a) for
the copy of $H$ which is between $u$ and $v$ there exists a cycle
in $D_c(H)$. Hence $\chi_c(G)=4$.
\end{proof}

Now we prove that it is \NP-complete to verify that the difference between chromatic
number and circular chromatic number of a given graph is greater
than or equal to $\frac{1}{k}$, when $k\ge 2$ is an arbitrary positive
integer is \NP-complete. Let $K$ be a graph with vertex set
$\{a,b,v_1,\ldots,v_{n-1} \}$ in which each $v_i$ is adjacent to 
every other $v_j$, $a$
is adjacent to $v_1, \ldots, v_{n-2}$, and $b$ is adjacent to
$v_{n-1}$.

\begin{lemma}
\label{kk} For all integers $0 \le x,y \le kn-1$, $K$ has a
$(kn-1,k)$-coloring $c$ with $c(a)=x$ and $c(b)=y$ if and only if
$x \neq y$.
\end{lemma}
\begin{proof}
If $x=y$, then a $(kn-1,k)$-coloring of $K$ can be transformed to a
$(kn-1,k)$-coloring of $K_n$ by identifying $a$ and $b$. And this
is impossible because $\chi_c(K_n)=n$. If $x \neq y$ without loss
of generality we can assume that $x=0$ and $ 0 < y \le
\frac{kn-1}{2}.$

First suppose that $ y \ge k$. In this case define a desired
$(kn-1,k)$-coloring $c$ by $c(a)=0$, $c(b)=y$, $c(v_i)=ik$ for $1
\le i \le n-2$ and $c(v_{n-1})=0$ .

Next suppose that $y < k$.  In this case define a desired
$(kn-1,k)$-coloring $c$ by $c(a)=0$, $c(b)=y$, $c(v_i)=ik$ for $1
\le i \le n-2$, and $c(v_{n-1})=y-k$.
\end{proof}

\begin{theorem}
For all positive integers $k \ge 2$ and $n \ge 3$, the following
problem is \NP-complete. A graph $G$ is given where $\chi(G)=n$,
and it is asked whether $\chi_c(G) \le n - \frac{1}{k}$?
\end{theorem}
\begin{proof}
Clearly, the problem is in \NP. We reduce {\sc Vertex Coloring} to
this problem. Consider a graph $G'$ as an instance of {\sc Vertex
Coloring}. It is asked whether the vertices of $G'$ can be colored
with $kn-1$ colors. We construct a new graph $G$ with the
property that $\chi_c(G) \le n - \frac{1}{k}$ if and only if the
vertices of $G'$ can be colored with $kn-1$ colors.

Construct a graph $G$ by replacing every edge $uv$ of $G'
\sqcup K_n$, the disjoint union of $G'$ and a copy of $K_n$,
 by $(K;a,b)$. Obviously, $\chi(G) \le n$. Since in every $(n-1)$-coloring 
 of $K$
 the vertices $a$ and $b$ must have different colors, thus $\chi(G)=n$.
  We know that $\chi_c(G) \le
n - \frac{1}{k}$ if and only if there exists a $(kn-1,k)$-coloring
$c$ for $G$.

First suppose that $\chi(G') \le kn-1$, and $c$ is a
$(kn-1)$-coloring of $G' \sqcup K_n$. For all copies of $K$ in
$G$, we have $c(a) \neq c(b)$. By Lemma~\ref{kk}, $c$ can be
extended to a $(kn-1,k)$-coloring of $G$. Thus $\chi_c(G) \le
n-\frac{1}{k}$.

Next suppose that $\chi(G') > kn-1$ and $c$ is a
$(kn-1,k)$-coloring of $G$. There exist two adjacent vertices $u$
and $v$ in $G'$ such that $c(u)=c(v)$. But by Lemma~\ref{kk}, the
copy of $K$ between $u$ and $v$ has no $(kn-1,k)$-coloring.
This is a contradiction. Thus $\chi_c(G) > n-\frac{1}{k}$.

\end{proof}

\section*{Acknowledgements}
The authors wish to thank Hossein Hajiabolhassan who drew their
attention to this subject, and for suggesting the problem.

\bibliographystyle{plain}
\bibliography{circul}

\begin{thebibliography}{1}

\bibitem{Guichard}
D.R. Guichard.
\newblock Acyclic graph coloring and the complexity of the star chromatic
  number.
\newblock {\em J. Graph Theory}, 17:129--134, 1993.

\bibitem{KLS}
S.~Khanna, N.~Linial, and S.~Safra.
\newblock On the hardness of approximating the chromatic number.
\newblock In {\em Proc. 2nd Israel Symp. on Theory of Computing and Systems},
  pages 250--260, 1993.

\bibitem{Vince}
A.~Vince.
\newblock Star chromatic number.
\newblock {\em J. Graph Theory}, 12:551--559, 1988.

\bibitem{West}
D.B. West.
\newblock {\em Introduction to Graph Theory}.
\newblock Prentice-Hall, Inc, United States of America, 2001.
\newblock 2nd Edition.

\bibitem{Zhu}
X.~Zhu.
\newblock Circular chromatic number: a survey.
\newblock {\em Discrete Math.}, 229:371--410, 2001.

\end{thebibliography}
\end{document}